\documentclass[11pt]{article}
\usepackage{amssymb} \usepackage{amsmath} \usepackage{graphicx}
\usepackage{epsfig,latexsym}

\begin{document}

\def\bef{\begin{figure}}
\def\eef{\end{figure}}
\newcommand{\ans}{ansatz }
\newcommand{\be}[1]{\begin{equation}\label{#1}}
\newcommand{\beq}{\begin{equation}}
\newcommand{\ee}{\end{equation}}
\newcommand{\beqn}[1]{\begin{eqnarray}\label{#1}}
\newcommand{\eeqn}{\end{eqnarray}}
\newcommand{\bd}{\begin{displaymath}}
\newcommand{\ed}{\end{displaymath}}
\newcommand{\mat}[4]{\left(\begin{array}{cc}{#1}&{#2}\\{#3}&{#4}
\end{array}\right)}
\newcommand{\matr}[9]{\left(\begin{array}{ccc}{#1}&{#2}&{#3}\\
{#4}&{#5}&{#6}\\{#7}&{#8}&{#9}\end{array}\right)}
\newcommand{\matrr}[6]{\left(\begin{array}{cc}{#1}&{#2}\\
{#3}&{#4}\\{#5}&{#6}\end{array}\right)}
\newcommand{\cvb}[3]{#1^{#2}_{#3}}
\def\lsim{\raise0.3ex\hbox{$\;<$\kern-0.75em\raise-1.1ex
e\hbox{$\sim\;$}}}
\def\gsim{\raise0.3ex\hbox{$\;>$\kern-0.75em\raise-1.1ex
\hbox{$\sim\;$}}}
\def\abs#1{\left| #1\right|}
\def\simlt{\mathrel{\lower2.5pt\vbox{\lineskip=0pt\baselineskip=0pt
           \hbox{$<$}\hbox{$\sim$}}}}
\def\simgt{\mathrel{\lower2.5pt\vbox{\lineskip=0pt\baselineskip=0pt
           \hbox{$>$}\hbox{$\sim$}}}}
\def\unity{{\hbox{1\kern-.8mm l}}}
\newcommand{\eps}{\varepsilon}
\def\ep{\epsilon}
\def\ga{\gamma}
\def\Ga{\Gamma}
\def\om{\omega}
\def\OM{\Omega}
\def\la{\lambda}
\def\La{\Lambda}
\def\al{\alpha}
\newcommand{\ov}{\overline}
\renewcommand{\to}{\rightarrow}
\renewcommand{\vec}[1]{\mbox{\boldmath$#1$}}
\def\tm{{\widetilde{m}}}
\def\mcirc{{\stackrel{o}{m}}}
\newcommand{\dm}{\delta m}
\newcommand{\tanb}{\tan\beta}
\newcommand{\nbar}{\tilde{n}}

\newcommand{\tnn}{\tau}	
\newcommand{\tnneff}{\tnn_\mathrm{eff}}	
\newcommand{\thetaeff}{\theta_\mathrm{eff}}

%

\newcommand{\Dsusy}{{susy \hspace{-9.4pt} \slash}\;}
\newcommand{\DCP}{{CP \hspace{-7.4pt} \slash}\;}
\newcommand{\mc}{\mathcal}
\newcommand{\gr}{\mathbf}
\renewcommand{\to}{\rightarrow}
\newcommand{\gtc}{\mathfrak}  
\newcommand{\wh}{\widehat}
\newcommand{\br}{\langle}
\newcommand{\kt}{\rangle}


\def\lsim{\mathrel{\mathop  {\hbox{\lower0.5ex\hbox{$\sim$}
\kern-0.8em\lower-0.7ex\hbox{$<$}}}}}
\def\gsim{\mathrel{\mathop  {\hbox{\lower0.5ex\hbox{$\sim$}
\kern-0.8em\lower-0.7ex\hbox{$>$}}}}}

\def\nn{\\  \nonumber}
\def\de{\partial}
\def\brf{{\mathbf f}}
\def\bbf{\bar{\bf f}}
\def\bF{{\bf F}}
\def\bbF{\bar{\bf F}}
\def\bA{{\mathbf A}}
\def\bB{{\mathbf B}}
\def\bG{{\mathbf G}}
\def\bI{{\mathbf I}}
\def\bM{{\mathbf M}}
\def\bY{{\mathbf Y}}
\def\bX{{\mathbf X}}
\def\bS{{\mathbf S}}
\def\bb{{\mathbf b}}
\def\bh{{\mathbf h}}
\def\bg{{\mathbf g}}
\def\bla{{\mathbf \la}}  
\def\bmu{\mathbf m }
\def\by{{\mathbf y}}
\def\bmu{\mbox{\boldmath $\mu$} }
\def\bunity{{\mathbf 1}}
\def\cA{{\cal A}}
\def\cB{{\cal B}}
\def\cC{{\cal C}}
\def\cD{{\cal D}}
\def\cF{{\cal F}}
\def\cG{{\cal G}}
\def\cH{{\cal H}}
\def\cI{{\cal I}}
\def\cL{{\cal L}}
\def\cN{{\cal N}} 
\def\cM{{\cal M}}
\def\cO{{\cal O}}
\def\cR{{\cal R}}
\def\cS{{\cal S}}   
\def\cT{{\cal T}}   
%


 \begin{titlepage}   



 \begin{center}   
 {\Large \bf Fast Neutron -- Mirror Neutron Oscillation and  \\
\vspace{3mm} 
Ultra High Energy Cosmic Rays }
 \end{center}

 \vspace{0.3cm} 

 \begin{center}   
 {\large Zurab Berezhiani$^{a,}$\footnote{E-mail: berezhiani@fe.infn.it, zurab.berezhiani@aquila.infn.it }
 and Lu\'{\i}s Bento$^{b,}$\footnote{E-mail: lbento@cii.fc.ul.pt
} } \\
 \end{center}

 \begin{center} \vskip 0.5 truecm 
 \centerline
 {$^a$\it Dipartimento di Fisica,
 Universit\`a di L'Aquila, I-67010 Coppito, AQ, Italy }
 \centerline{\it and INFN, Laboratori Nazionali del Gran Sasso,
 I-67010 Assergi, AQ, Italy}
 \vskip 0.2 truecm
 \centerline{$^b$\it Universidade de Lisboa, Faculdade de Ci\^ encias }
 \centerline{ \it Centro de F\'{\i}sica Nuclear da Universidade de Lisboa}
  \centerline{ \it Avenida Prof.\ Gama Pinto 2, 1649-003 Lisboa, Portugal }
 \end{center}

\begin{abstract}

If there exists the mirror world, a parallel hidden sector of particles 
with exactly the same microphysics as that of the observable particles, 
then the primordial nucleosynthesis constraints require that 
the temperature of the cosmic background  of mirror relic photons 
should be smaller than that of the ordinary relic photons, $T'/T < 0.5$ or so. 
On the other hand, the present experimental and astrophysical limits 
allow a rather fast neutron - mirror neutron oscillation in vacuum, 
with an oscillation time  $\tnn \sim 1$ s, 
much smaller than the neutron lifetime. 
We show that this could provide a very efficient 
mechanism for transporting ultra high energy protons 
at large cosmological distances. 
The mechanism operates as follows: 
a super-GZK energy proton scatters a relic photon producing a neutron
that oscillates into a mirror neutron which then decays into a
mirror proton.
The latter undergoes a symmetric process, scattering a mirror relic photon 
and producing back an ordinary nucleon, but only after traveling 
a distance $(T/T')^{3}$ times larger than ordinary protons.
This may relax or completely remove the GZK-cutoff 
in the cosmic ray spectrum and also explain the correlation 
between the observed ultra high energy protons and 
far distant sources as are the BL Lacs.     

\end{abstract}

 \end{titlepage}



\section{Introduction}

The hypothesis that there may exist a mirror world,
a hidden parallel sector of particles 
which is an exact duplicate of our observable world, 
has attracted a significant interest over the past years 
in view of its implications for particle physics
and cosmology \cite{Mirror}-\cite{M-cosm} 
(for reviews, see~\cite{IJMPA-F,IJMPA-B}.) 
The basic concept can be formulated as follows: 
one has a theory given by the
product $G\times G'$ of two identical gauge factors with
identical particle contents, 
where the ordinary particles belong to $G$ 
and are singlets of $G'$, and vice versa, mirror particles 
belong to $G'$ and are singlets of $G$.
(From now on all fields and quantities of the
mirror (M) sector will be marked with $'$ to distinguish from
the ones of the ordinary/observable (O) world.)
The Lagrangians of both sectors are identical i.e., all coupling
constants (gauge, Yukawa, Higgs) have the same pattern
in the O- and M-worlds,
which means that there is a discrete symmetry,
the so called mirror parity, under the interchange of 
$G$ and $G'$
gauge and matter fields~\cite{FLV}.
The two worlds may be considered as parallel branes embedded
in a higher dimensional space,
with O-particles localized on one brane and the M-ones on another brane 
while gravity propagates in the bulk, a situation that can emerge
e.g.\ in the context of $E_8\times E'_8$ superstring theory.

If the mirror sector exists, then the Universe 
should contain along with the ordinary photons, electrons, nucleons 
etc., also their mirror partners with exactly the same masses.  
However, the fact that the O- and M-sectors have the same microphysics, 
does not imply that their 
cosmological evolutions should be the same too.
If mirror particles had the same cosmological abundances 
in the early universe as ordinary ones, this would be in immediate 
conflict  with Big Bang Nucleosynthesis (BBN); 
their contribution would increase the Hubble expansion rate 
by a factor of $\sqrt2$ which is equivalent to an effective number 
of extra neutrinos $\Delta N_\nu \simeq 6.14$. 
Thus, at the BBN epoch the temperature of M-sector must be 
smaller than the temperature of O-sector.  
In this view, one can accept the following paradigm~\cite{BCV,IJMPA-B}:
at the end of inflation the two systems are (re)heated 
in an non symmetric way which can be natural in the context 
of certain inflationary models: namely, 
the M-sector gets a lower temperature than the O-sector.  
The two systems evolve further adiabatically,
without significant entropy production,
and interact with each other so weakly that the equilibrium
between them cannot be achieved in subsequent epochs.
In this way, the ratio of their temperatures
would be kept nearly constant during the expansion of the universe, 
until the present days. 

The parameter $x=T^\prime/T$, with $T$ and $T'$ being the 
temperatures of the relic CMB photons in O- and M-sectors respectively, 
will play a key role in our further considerations. 
It is related to the effective value  $\Delta N_\nu$ at 
the BBN epoch as 
$\Delta N_\nu \simeq 6.14\cdot  x^{4}$ 
\cite{BDM,BCV}. 
Hence, by taking a conservative bound 
$\Delta N_\nu < 1$ we get an upper limit $x < 0.64$, 
while for e.g. $x = 0.5 ~ (0.3)$ one would have respectively 
$\Delta N_\nu \simeq 0.38 ~(0.05)$. 

Due to the temperature difference between the two worlds, 
all the key epochs in the mirror sector, baryogenesis, nucleosynthesis,  
recombination, etc., proceed at somewhat different conditions than in 
the observable sector of the Universe~\cite{BCV}. 
In particular, in certain baryogenesis scenarios the M-sector gets 
a larger baryon asymmetry than the O-sector, 
and it is plausible that the mirror baryon density $\Omega'_B$ is 
comparable with the ordinary  baryon density $\Omega_B\simeq 0.04$. 
This situation emerges in a particularly appealing way in the leptogenesis 
scenario due to lepton number leaking from the O-sector to M-sector
which implies that $\Omega'_B /\Omega_B\geq 1$ \cite{BB-PRL,IJMPA-B}.  
Hence, mirror matter, being invisible in terms of ordinary photons, 
but interacting with ordinary matter via gravity, could 
be a viable dark matter candidate with specific observational 
signatures~\cite{BCV,M-cosm,IJMPA-B}.

Besides gravity, the two sectors could communicate by other means.
There can exist interactions between the O- and M-fields mediated by some messengers, which may be pure gauge singlets or some fields in mixed 
representations of $G\times G'$, an axion from a common Peccei-Quinn 
symmetry, as well as  extra gauge bosons acting with both sectors, 
related e.g.\ with a common  flavor or $B-L$ gauge 
symmetries~\cite{PLB98,IJMPA-B}. 
Such interactions could induce the mixing of some neutral O-particles, 
elementary as well as composite,  with their M-counterparts.
In particular, photons could have a kinetic mixing with
mirror photons~\cite{Holdom,FLV}, which can be searched in the 
ortho-positronium oscillation into mirror ortho-positronium \cite{ortho} 
and can also be tested with dark matter detectors~\cite{IJMPA-F},  
neutrinos could mix with mirror neutrinos, with interesting 
astrophysical implications~\cite{FV},
neutral pions and/or Kaons could mix with their mirror partners~\cite{IJMPA-B}, etc.   


In this paper we explore the mixing between the neutron $n$ 
and the mirror neutron $n'$ due to a small mass mixing term 
$\dm \,(\ov{n} n' + \ov{n}' n)$.  As we have shown in ref.~\cite{BB-nn},
the experimental and astrophysical limits do not exclude 
a rather rapid $n -n'$ oscillation, with a timescale $\tnn$ 
much smaller than the neutron lifetime $\tau_n \simeq 10^3$ s, 
and in fact as small as $\tnn =1/\dm\sim 1$ s.  
Such an intriguing possibility, apart from the fact that
can be tested in small scale "table-top" experiments,
can also have far going astrophysical implications \cite{BB-nn}
(see also~\cite{Nasri}). 
Namely, 
we shall show below that it could provide a very
efficient mechanism of transport of ultra high energy protons  
at large cosmological distances and thus,
could explain rather naturally the excess of events 
above the Greisen-Zatsepin-Kuzmin (GZK) cutoff~\cite{GZK}.   

The paper is organized as follows. In section 2 we briefly discuss 
the physics of $n-n'$ mixing and the experimental 
limits on  $n-n'$ oscillation. 
Then we discuss the implications for the cosmic rays at
super-GZK energies (sect. 3) and conclude with section 4.  


\section{Mirror sector and neutron - mirror neutron mixing} 

At a fundamental level, the neutron - mirror neutron mixing should result 
from some effective couplings between quarks and mirror quarks.
The situation we are interested in, that the two particles, 
neutron and mirror neutron, are exactly degenerate in mass 
when their mixing is turned off,
is naturaly obtained if the interchange between ordinary and mirror fields
-- the mirror parity -- is an exact symmetry of the Lagrangian.

Consider, for example, 
that both ordinary and mirror sectors are described by 
the minimal gauge symmetry group i.e.,
the Standard Model gauge group
$G = SU(3) \times SU(2) \times U(1)$ 
applies to the O-sector comprising
the left-handed quarks and leptons\footnote{   
For convenience, we omit the symbols L and R 
as well as the internal gauge, spinor and family indices. 
Antiparticles will be termed as $\tilde{q}$, $\tilde{l}$, etc..}
$q = (u, d)$, $l = (\nu, e)$,
right-handed iso-singlets
$u$, $d$, $e$,
and the yet unobserved Higgs $\phi$,
whereas the duplicate gauge symmetry group
$G' = SU(3)' \times  SU(2)'\times U(1)'$
describes the hidden mirror sector consisting of mirror quarks, leptons and Higgs:
left-handed $q' = (u', d')$, $l' = (\nu', e')$, 
right-handed $u'$, $d'$, $e'$, 
and mirror Higgs $\phi'$.
Mirror parity forces that all coupling constants, gauge, Yukawa and Higgs,
are identical for both sectors, 
the mirror Higgs 
vacuum expectation value (VEV) 
$\langle \phi' \rangle = (0,v')$ is identical 
to the standard VEV $\langle \phi \rangle = (0,v)$,  i.e.\ $v'=v$,
and hence the mass spectrum of mirror particles,
elementary as well as composite, 
is exactly the same as that of ordinary ones.

As usual, we assign a global lepton charge $L = 1$ to leptons and 
a baryon charge $B = 1/3$ to quarks.
Similarly, we assign a lepton charge $L'=1$ to mirror
leptons and a baryon charge $B'=1/3$ to mirror quarks.
By definition, ordinary fermions have $L'=B'=0$
and the mirror particles have null $B$ and $L$ charges.
If $L$ ($L'$) and $B$ ($B'$) were exactly conserved the 
neutrinos $\nu$ ($\nu'$) would remain massless and phenomena like 
proton (mirror proton) decay or 
neutron - antineutron $n - \nbar $ ($n' - \nbar' $) oscillation \cite{nnbar}
would be impossible.

The neutron - mirror neutron mixing we are interested in violates 
both baryon numbers $B$ and $B'$ but conserve the combined quantum number $\bar{B} =B-B'$.\footnote{ 
For definiteness, we identify both $n$ and $n'$ as states that have 
$\bar{B} =1$, though, in accordance with our definitions, 
the three constituents of $n'$ should be identified as "mirror antiquarks". 
As far as we have not yet any relations with the mirror 
world, this should have no political consequences.
} 			
It is possible to generate such mixing by incorporating couplings between 
three quarks and three mirror quarks given by the D=9 operators~\cite{BB-nn}
\be{n-npr}
\cO_9^{\rm mix} \sim 
\frac{1}{\cM^5} (udd)(u'd'd')  +  \frac{1}{\cM^5} (qqd)(q'q'd')  
 + {\rm h.c.}  \;.
\ee
If one postulates $\bar{B}$ conservation, the operators 
$\cO_9 \sim (udd)^2 /M^5$ or $(qqd)^2 /M^5$ 
relevant for neutron - antineutron oscillation ($\Delta \bar{B} =2$) are forbidden,
whereas the operators $\cO_9^{\rm mix}$ and 
the neutron - mirror neutron oscillation ($\Delta \bar{B} =0$) 
are allowed.\footnote{The possible      
theoretical models for the generation of the operators (\ref{n-npr}) were discussed 
in refs.~\cite{BB-nn,Nasri}. } 

A similar situation holds for the lepton sector.
The usual D=5 operator 
$\cO_5 \sim(l\phi)^2 /M$ 
and the resulting, after symmetry breaking, small Majorana masses for neutrinos, 
$m_\nu \sim v^2/M$ \cite{Weinberg},
violate the quantum number $\bar{L}= L - L'$ by two units.
However, the operator
\be{l-lpr}
\cO^{\rm mix}_5 \sim  \frac{1}{M} (l \phi)(l' \phi')  
+ {\rm h.c.} \;   
\ee
mixes ordinary and mirror neutrinos~\cite{FV}, but conserves $\bar{L}$.
If one postulates $\bar{L}$ conservation, then
one can obtain light Dirac neutrinos from $\cO^{\rm mix}$ \cite{BB-nn}
with ($\Delta {L} = \Delta {L'} =1$) 
mass terms $m_{\nu} \sim v^2/M$,   
where the right-handed components are mirror anti-neutrinos
 $\nu'_R = {{\nu}'_L}^c$.
These Dirac masses and mixings can explain the neutrino flavor oscillation phenomena equaly well as pure $\Delta {L} = 2$ Majorana masses 
but distinguish from them by the absence of 
neutrinoless $2\beta$ decay.

As stated above the operators (\ref{n-npr}) generate 
a mass mixing term between $n$ and $n'$, 
$\dm \,(\ov{n} n' + \ov{n}' n)$.
Taking into account that the matrix elements of the operators 
$\cO_9^{\rm mix}$ 
between the neutron states are typically of the order of
$\Lambda_{\rm QCD}^6 \sim 10^{-4}$ GeV$^6$,
one estimates 
\be{deltam}
\dm \sim \left(\frac{10 \, {\rm TeV} }{\cM}\right)^5 \cdot 
10^{-15} \, {\rm eV}	\; .
\ee
If neutrons and mirror neutrons were completely free particles 
they would oscillate with maximal mixing angle and a caracteristic 
oscillation time equal to
$\tnn = 1/\dm=  0.66\cdot (10^{-15} ~ {\rm eV}/\dm)$ s.
Remarkably, the present available experimental data does allow an 
oscillation time as small as $\tnn \sim 1$ s  or so~\cite{BB-nn},  
much smaller than the bound on the neutron - antineutron oscillation time
$\tau_{n\nbar}=1/\dm_{n\nbar} > 10^8$ s, 
and even shorter than the neutron lifetime itself, $\tau_n \approx 10^3$ s.
The reason for such a weak bound is twofold: 
i) contrary to the antineutron case the mirror neutron is an invisible particle 
and therefore the $n-n'$ oscillation can be experimentally observed only 
as neutron disappearance (or regeneration);
ii) as a rule, neutrons are not actualy free particles, but are rather subject 
to external electromagnetic or strong interactions.

The evolution of free non-relativistic neutrons is described
by the effective Hamiltonian in $n-n'$ space,
\be{Hn-npr}
H = \mat{{p^2}/{2 m} -i \,{\Gamma_n}/{2} - V}{\dm} 
{\dm} {{p^2}/{2 m} -i \,{\Gamma_n}/{2}  - V^\prime}
	\;.
\ee
The neutron mass $m$ and decay width $\Gamma_n =1/\tau_n$
are precisely the same for mirror neutrons, due to exact mirror parity,
but the potentials $V$ and $V'$ felt by $n$ and $n'$ 
are not quite the same. Then, for $\Delta V = | V-V' | \neq 0$,  
the effective oscillation time $\tnneff$ and 
effective mixing angle $\thetaeff $ become  
\beqn{thetaeff}
\tnneff = \tau / \sqrt{1+ (\Delta V/2\, \dm)^2 }  \, ,  \nonumber \\
\sin 2\, \thetaeff = \tnneff \, \dm = \tnneff /\tau  \; , 
\eeqn
and thus the transition probability from an initial pure neutron state at 
$t=0$ to a mirror neutron state, or vice versa from $n'$ to $n$, is  
\be{Pn-npr}
P_{nn'}(t) = 
(\tnneff /\tau)^2  \, \sin^2 (t/\tnneff) \, e^{-t/\tau_{n}}  \; . 
\ee

Neutrons propagating in normal matter undergo a large effective potential 
$V\sim 10^{-7}$ eV, 
which makes the effective mixing angle, 
$\thetaeff \approx \dm/V$,
negligibly small and blocks the oscillation.
But even in experiments with free neutrons in vacuum,
the terrestrial magnetic field $B\simeq 0.5$ G induces a small 
but significant potential
$V\simeq \mu B\simeq 3\cdot 10^{-12}$ eV,
where $\mu \approx 6\cdot 10^{-12}$ eV/G 
is the neutron magnetic moment. 
On the other hand, $V'$ is vanishingly small since no significant 
mirror magnetic fields are expected on Earth.
Then, for $\dm < V$,  
the effective oscilation time is   
$\tnneff \approx 2/V \simeq 4.4 \cdot 10^{-4}$ s
and the average transition probability becomes 
$P_{nn'} \simeq 2\, (\dm/V)^2$.
For e.g.\
$\dm =10^{-15}$ eV it amounts to 
$P_{nn'} \approx 2\cdot 10^{-7}$,
which is a marginal effect indeed.
Hence, in order to have a sizeable oscillation 
the magnetic field must be shielded.

In the ILL-Grenoble experiment designed to search for
neutron - antineutron oscillation, the magnetic field 
was reduced to $B\sim 10^{-4}$ G \cite{Grenoble},
giving $V\simeq \mu B\sim 6\cdot 10^{-16}$ eV. 
Cold neutrons propagated in vacuum with an average speed of $600$ m/s 
and effective time of flight $t \simeq 0.1$ s, 
in a mu-metal vessel shielding the magnetic field.
No antineutrons were detected and the limit
$\tau_{n\nbar} > 0.86 \cdot 10^8$ s was reported.
On the other hand, it was observed that about 5$\%$ of the 
neutrons disappeared. 
Most of the losses can be attributed to scatterings with the 
walls in the drift vessel, but a fraction of them might be due to the 
oscillation into an invisible particle, the mirror neutron.
For a vacuum oscillation time $\tnn = 1$ s, the mixing mass 
$\dm = 1/\tnn \approx 6\cdot 10^{-16}$ eV 
matches the magnetic potential energy and 
$\tnneff \approx \tnn = 1$s.
Then, during the small time of flight $t \simeq 0.1$ s, 
a fraction of the neutrons equal to
$P_{nn'}(t) \approx (t/\tnn)^2 \approx 1 \% $
would have oscillated into a mirror neutron state 
contributing so to the observed neutron beam deficit.
(If the whole deficit was entirely due to 
$n-n'$ oscillation, one would have $\tnn \simeq 0.4$ s.) 
This is clearly a speculation but the remarkable fact is, 
it cannot be excluded from the observations made so far.
Second, the oscillation time scale may be much smaller than the neutron lifetime itself.
Dedicated experiments are needed to put stronger bounds on such phenomenum.

Another potential source of evidence or exclusion of $n - n'$ oscillation 
could come from nuclear physics.
In the case of neutron - antineutron oscillation the observed nuclear stability has provided the strongest bounds~\cite{Chetyrkin}.
That is not the case of $n - n'$ oscillation.
Suppose that a nucleus $(A,Z)$ may decay into the isotope $(A-1,Z)$ 
emitting an invisible mirror neutron $n'$ and/or its beta decay products 
$p' e' \tilde{\nu}'_e$.
But then, also the decay with neutron and/or 
$p\, e\, \tilde{\nu}_e$  emission
is kinematicaly allowed because these particles are 
exactly degenerate in mass with their mirror counterparts.
We have confirmed by inspection that all nuclear ground states 
satisfying those kinematic conditions suffer from standard nuclear instability and have short lifetimes~\cite{ntables}.
In such a case, the mirror channels are expected to be invisible, 
overwhelmed by the standard processes, because the effective $n - n'$ 
mixing angle in nuclei is vastly suppressed by the neutron nuclear energy.
For $\dm = 10^{-15}$ eV and a potential energy as small as $V = 1$ keV, 
one gets $\thetaeff^2 \approx 10^{-36}$.
Similar arguments apply to the stability of astrophysical objects like neutron stars. 
Hence, the only realistic limit that remains is $\tnn > 1$ s or so,
indirectly extracted from the neutron beam controlling
procedure at the ILL-Grenoble reactor experiment. 

Now we show that the possibility of such a fast oscillation 
opens up very intriguing prospects for understanding 
the problems concerning the cosmic rays at ultra high energies (UHE). 


\section{Implications for ultra high energy cosmic rays } 

It was pointed out a long time ago \cite{GZK} that the cosmic microwave 
background (CMB) of relic photons makes the universe opaque 
to the ultra high energy cosmic rays (UHECR). 
In particular, protons with energies 
above the $p\gamma \to N\pi$ reaction threshold,   
\be{Eth}
E_{\rm th} \approx \frac{m_\pi m_N}{2\, \eps_\gamma} \simeq  
6 \cdot 10^{19} \; {\rm eV} 
	\;, 
\ee  
with $m_\pi$ and $m_N$ being respectively the pion and nucleon masses 
and $\eps_\gamma \sim 10^{-3}$ eV a typical relic photon energy,
cannot propagate at large cosmological distances 
without losing their energy. 
As a result, one expects to see an abrupt cutoff in the 
spectrum of cosmic rays above the threshold energy $E_{\rm th}$, 
the so-called GZK-cutoff or GZK-feature~\cite{Review}. 
 
For energies $E > E_{\rm th}$, \
the proton mean free path (m.f.p.) with respect to $p\gamma \to N\pi$ 
scattering, is roughly 
$l_p \simeq (\sigma n_\gamma)^{-1} \simeq 5 \; {\rm Mpc}$, 
where $n_\gamma \approx 400$ cm$^{-3}$ is the number density 
of CMB photons and $\sigma \simeq 0.1$ mb is the characteristic  cross section.  
On average, protons lose per scattering a fraction 
$y\simeq 0.2$  of their energy.\footnote{
In reality, both $\sigma$ and $y$ 
have some energy dependence which will be neglected in our qualitative 
estimations for the sake of simplicity.  
We also neglect the energy losses due to $e^+ e^-$ pair production, 
and apologize to the super-GZK super-experts for such 
a naive treatment of the problem.
} 
Therefore, super-GZK protons should lose their energy when  
traveling at large cosmological distances.
Namely, while traveling a distance $R$ from the source to the Earth, 
a proton with initial energy $E_R \gg E_{\rm th}$ at the source, 
will suffer on average $k_R=R/l_p$ collisions $p\gamma\to N\pi$
and so its energy will be reduced to $E \simeq E_R (1-y)^{k_R}$.
(Energy losses stop when the proton energy gets 
down below the threshold energy $E_{\rm th}$ and the effective proton 
m.f.p.\ becomes very large.) 
Therefore, the distance to the source of a proton that was emitted 
with initial energy $E_R$ and is detected at the Earth with a super-GZK 
energy $E > E_{\rm th}$ can be roughly estimated as\footnote{
In fact, the UHE cosmic rays travel long distances 
transforming continuously from protons to neutrons and back. 
The  $p\, \gamma \to N \, \pi$ scatterings produce both protons and neutrons 
($p\, \pi^0$ or $n\, \pi^+$) with nearly equal probabilities (isotopical symmetry). 
If the Lorentz factor $\gamma =E/m_N$ is not too large, 
neutrons suffer $\beta$-decay $n\to p\, e\, \tilde{\nu}_e$, 
after traveling a distance of about
$l_{\rm dec}= \gamma\, c\, \tau_n \simeq  (E/10^{20} \, {\rm eV})$ Mpc,  
transferring the whole energy back to the proton. 
Therefore, at energies below $5\cdot 10^{20}$ eV the cosmic ray carriers 
will be mostly protons, while for $E > 5\cdot 10^{20}$ eV 
they will be both protons and neutrons in nearly equal proportions. 
Anyway, the existence of neutrons does not change the propagation distance, 
since their scatterings off the CMB photons,
$n\, \gamma \to p\, \pi^-(n\, \pi^0)$, 
have nearly the same cross sections as protons, and thus 
the neutron m.f.p.\ $l_n$ is also $\sim 5$ Mpc. 
} 
\be{R}
R \sim k_R \, l_p \sim \log ({E_R}/{E}) \cdot  50 \; {\rm Mpc}
	\;.
\ee  
For example, the proton with energy of $3\cdot 10^{20}$ eV detected 
by Fly's Eye had to have an initial energy of about 
$E_R \simeq 3\cdot 10^{22}$ eV, if it came from a distance $R=100$ Mpc 
(and in fact, no astrophysical source is seen at this direction within the 
radius of 100 Mpc).
However, for $R > 400$ Mpc one would need an initial energy 
$E_R > 3\cdot 10^{28}$ eV, bigger than the Planck Energy! 


As for the hypotetical UHE mirror protons, they can propagate 
much larger distances through the background of relic mirror photons  
as far as the mirror CMB has a smaller temperature $T' = x\,T$, 
with $x < 0.5$ as required by the BBN bounds. 
First, because the typical relic photon energy rescales as 
$\eps'_\gamma=x \, \eps_\gamma$ 
and second, because the mirror photon density is suppressed
by a factor of $x^{3}$, $n'_\gamma = x^3 n_\gamma$, 
while the scattering processes and their cross sections 
are the same in both sectors.
As a result, the threshold energy is somewhat higher for  mirror protons 
while their  m.f.p.\ is drastically amplified: 
\be{Ethpr}
E'_{\rm th} \simeq x^{-1} E_{\rm th} \;,   \qquad
l'_p \simeq \, x^{-3} l_p
	\; . 
\ee
Thus, the mirror UHE protons can travel large cosmological distances 
without significant loss of energy.  
With decreasing $x$ values,  the GZK-cutoff in the mirror 
UHECR spectrum not only shifts to higher energies 
with respect to the ordinary GZK-cutoff, but also becomes less sharp. 
In particular, for $x=0.1$ one gets 
$l'_p \simeq 5 \cdot 10^{3}$ Mpc, 
of the order of the present Hubble radius, 
which means that  mirror protons would cross the universe 
without losing energy and in this case the GZK-feature 
would simply not exist in the mirror proton spectrum. 

Let us discuss now what can happen if there exists  
$n-n'$ oscillation with $\tnn \ll \tau_n$. 
Such a proccess may dramatically change
the paradigm of cosmic ray propagation
by providing a mechanism that converts ordinary protons into mirror ones
and vice versa at super-GZK energies.  
Obviously, for relativistic neutrons both oscilation and decay 
lengths rescale with the Lorentz factor, hence 
$l_{\rm osc}/l_{\rm dec}=\tnn /\tau_n$.  
For say, $\tnn \sim 1$ s, the oscillation length 
of a neutron with an energy of $10^{20}$ eV is about 1 kpc, 
and so it oscillates many times into/from a mirror neutron 
before 
decaying or scattering a relic photon.

Consider a flux $J_R$ of ordinary protons with energy 
$E_R \gg E_{\rm th}$ emitted from a distant source. 
After traveling a path $\sim l_p\simeq 5$ Mpc each proton  
scatters once with a CMB photon producing  
$p \,\pi^0$ or $n\, \pi^+$, with a 1/2 probability for each channel.  
A neutron produced in this way oscillates,  
with a mean probability $w$, into a mirror neutron $n'$ 
which decays later into $p' e' \tilde{\nu}'$, while with a probability 
$1-w$ it survives as ordinary neutron which then decays 
into $p e \tilde{\nu}$.
In vacuum $n-n'$ are maximally mixed and thus $w=1/2$.   
However, external factors like cosmological magnetic fields (see below)  
may change  the effective mixing angle in the medium, 
and we leave $w$ as arbitrary,
keeping though in mind that our mechanism requires anyway large enough 
mean values of  $w$.\footnote{ 	
For simplicity, we also assume that 
neutrons decay (into $p$ or $p'$) before scattering 
off the CMB, thus "integrating out" the UHECR propagation at 
super-GZK energies in terms of $n-n'$ states, and effectively reducing 
it to the chain of continuous transformation from $p$ to $p'$ and 
vice versa. 
Strictly speaking, for $w=1/2$ this is valid  
for energies $E < 10^{21}$ eV, when the neutron decay length  
$l_{\rm dec}\simeq (E/10^{20}~ {\rm eV})$ Mpc 
is smaller than the m.f.p.\ of the oscillating $n-n'$ system with 
respect to the scattering off the ordinary CMB photons.    
The latter is about $2\, l_p\sim 10$ Mpc, 
since the mixed $n-n'$ state 
propagates half of the time as $n$ 
and half of the time as $n'$.}
 
Hence, after first scattering at a typical distance $\sim l_p$, 
a fraction $w/2$ of the initial proton flux $J_R$ 
will be converted into mirror protons and 
the fraction $1-w/2$ will remain in terms of ordinary protons, 
both fractions having a typical energy $(1-y) E_R$. 
The fraction of ordinary protons will  scatter CMB once again after a 
typical distance $\sim l_p$,  
and still produce, with a probability $w/2$, mirror neutron $n'$ which 
decays into $p'$, and so on for the subsequent scatterings.  
Therefore, after $k$ scatterings we observe 
not only energy dissipation, $E_k \simeq E_R (1-y)^k$, 
but also a deficit in the number of particles:  
the flux of ordinary nucleons  reduces to 
$J_k \simeq J_R (1-w/2)^k$.  
Namely, for $w=1/2$, 
about a half of the initial ordinary protons 
will be converted into mirror ones just after two scatterings.
For arbitrary $w$, the amount of collisions needed for the e-fold 
conversion $p\to p'$ can be estimated as $k_e=-1/\ln(1-w/2)$. 
However, mirror protons produced in this way, can be also converted back 
to ordinaries  by scattering mirror CMB photons. 

The subsequent story strongly depends on the value of the 
parameter $x$. There are three different regimes: 
small, moderate and large values of $x$. 

Let us start with the case of very small $x$, say  $x=0.1$. 
In this case the m.f.p.\ of mirror protons becomes 
of the order of the present Hubble radius, 
$l'_p = x^{-3} l_p \simeq 5 \cdot 10^{3}$ Mpc.  
Hence, mirror protons can pass the whole universe 
without scattering off the mirror CMB and thus never turn back into the form 
of ordinary protons. 
Hence, for very small $x$ the mirror sector acts like a sink where 
ordinary protons of the super-GZK energies fall and disappear. 
This can only make the cosmic ray spectrum above the threshold energy 
(\ref{Eth}) even more abrupt.  

In the other extreme case of very large $x$ ($\approx 1$), 
when the mirror m.f.p.\ $l'_p$ is comparable to $l_p$, 
the mirror protons transform back into ordinaries 
after scattering with mirror CMB exactly in the same
manner as described above, but one does not gain much.
For example, for $x=1$ the UHECR would suffer exactly 
the same energy losses in both $p$ and $p'$ forms and thus the effective 
propagation distance would remain as in eq.~(\ref{R}). 

The most interesting are the moderate values of $x$, 
in the interval $x= 0.3-0.5$ or so. 
In this case mirror protons, 
before scattering their CMB and transforming back to 
ordinary particles, can travel with m.f.p.\ $l'_p =x^{-3} l_p \sim 40-200$ Mpc, 
which is still much smaller than the Hubble radius. 
Then, the evolution of the system at the cosmological distances 
can be described as the propagation of a mixed $p-p'$ 
state that changes its nature from ordinary to mirror proton or vice versa, 
with a mean probability $w/2$ after each collision. 
Its propagation distance corresponding to $k_R$ scatterings  
becomes effectively  
for $k_R \leq 2k_e$, instead of eq.~(\ref{R}),
\be{Reff}
R_{\rm eff} \sim \frac{1}{2}k_R (l_p + l'_p)  \sim
\frac{1+x^{-3}}{2} \; \log\left(\frac{E_R}{E}\right) \cdot 50 \;\; {\rm Mpc}
	\;,
\ee  
which e.g.\ for $x\simeq 0.4$ gives  
$R_{\rm eff} \sim 400 \; {\rm Mpc} \cdot \log(E_R/E)$. 
Therefore, such a mixed UHECR can travel long cosmological 
distances 
without suffering substantial energy losses. 

On the other hand, for most of the time it travels dressed 
as $p'$, and the probability of finding it in the form of 
an ordinary proton (or neutron) 
at a large distance from the source is small: 
$W_p \sim l_p/l'_p \simeq x^3$, 
which is about 7\% for $x\simeq 0.4$. 
Hence, one loses a numerical factor of $x^3$ in the flux 
but still overcomes the strong exponential suppression. 
Observe also, that at large distances from the source, 
it does not matter which was the original state, $p$ or $p'$
-- the propagating state forgets its initial condition. 
So, if there are also sources of mirror UHECR in the universe  
with fluxes order of magnitude stronger than ordinary ones, 
and this is not unnatural if one considers mirror baryons as a constituent 
of dark matter, then their contribution may compensate the deficit 
created by the  reduction factor $W_p\simeq x^3$.\footnote{
For intermediate energies between 
$E_{\rm th}$ and $E'_{\rm th} \simeq E_{\rm th}/x$ 
there can be an interesting feature: 
ordinary protons can transform into mirror ones, 
but mirrors cannot efficiently transform back into ordinaries. 
Hence, a dip could emerge in the observed 
cosmic ray spectrum between $E_{\rm th}$ and $E'_{\rm th}$
that would be very prominent if the thresholds were ideally sharp.
But the integration over the thermal distribution of relic photons 
smooths out such a  dip.
}   

This discussion indicates that moderate values of $x$ 
around $0.3-0.5$ are preferable. 
In this case the mirror m.f.p.\ is of the order of 100 Mpc which 
consents a feasible correlation \cite{TT} between the UHECR
events observed by AGASA and the distant astrophysical 
sources known as BL Lacs, a special sort of 
blazars that could be plausible candidates for natural 
accelerators of protons to the ultra high energies. 
It seems also natural that BL Lacs, consisting 
essentially of central black holes surrounded by accreting matter, 
ordinary as well as mirror, could be accelerating sources 
for both ordinary and mirror protons: 
a black hole is a black hole, does not matter to which kind of matter. 



The following remark is in order. 
The $n-n'$ oscillation of UHE neutrons 
may be suppressed by cosmological magnetic fields. 
For slow neutrons they are unimportant, 
but they can suppress the oscillation of very fast neutrons  
because the transverse magnetic field is
amplified by the Lorentz factor $\gamma =E/m_N$. 
Namely, the evolution of fast neutrons in external magnetic    
fields is equivalent in the neutron rest frame 
to a Hamiltonian given by eq.~(\ref{Hn-npr}) with  
$V=\gamma \mu B_{\rm tr}$ and  $V'=\gamma \mu B'_{\rm tr}$, 
where $B_{\rm tr}$ and  $B'_{\rm tr}$ are respectively, the 
transverse components of ordinary and mirror magnetic fields. 
Therefore, for super-GZK cosmic rays, $E > 10^{20}$ eV ($\gamma >10^{11}$),
the effective mixing angle (\ref{thetaeff}) in a medium with arbitrary 
magnetic field orientations 
is effectively suppressed unless $B , B' < 10^{-15}$ G or so.

One can object whether this is realistic. 
The galactic magnetic fields measured by the Faraday rotation 
are rather large, of the order of $10^{-6}$ G, but they do not have 
much memory about the magnitude of primordial seeds as 
they are amplified many orders by the galactic dynamo. 
The magnetic fields observed in the central regions of clusters are also large. 
Magnetic fields could be also improperly big in filaments, 
however, cosmic rays travel mostly in voids where in principle the 
fields might be small enough, $\leq 10^{-15}$ G. 
In reality we do not have any direct observational 
indications about  the size of magnetic
fields at large, order 100 Mpc scales and moreover, there is no
established mechanism for their generation. 
The observational data only give an upper bound on the latter, 
roughly $B < 10^{-9}$ G or so.  
The physical mechanisms related to the dynamics of primordial plasma 
before recombination yield primordial magnetic seeds 
$B<10^{-19}$ G at the 100 Mpc scales~\cite{Dolgov},  
while $B'$ should be even smaller due to an earlier recombination 
of mirror matter, $z'_{\rm rec} \simeq x^{-1} z_{\rm rec}$~\cite{BCV}.  

On the other hand, numerical simulations \cite{dolag} indicate 
that the magnetic fields observed in clusters need primordial 
seeds $> 10^{-13}$ G at the scales order 10 Mpc,   
which seems difficult to reconcile with $B < 10^{-15}$ G  
in voids.  Observe however, that it is not needed at all that the
magnetic fields are small in every region: 
as far as a neutron with energy $E> 10^{20}$ eV travels
a distance larger than 1 Mpc before it decays, 
it is enough for the oscillation to occur that within this distance 
it can meet with a reasonable probability patches 
where $B_{\rm tr}$  is accidentally smaller than $10^{-15}$ G.  
It is not just the magnetic field intensity that matters, 
but it is also its orientation with respect to the neutron (mirror  neutron) 
direction of motion, given that the transverse magnetic field component is amplified by the Lorentz factor but the longitudinal one is not. 
Therefore,  the $n-n'$ oscillation is not be blocked also by  
a magnetic field as large as $B=10^{-14}$ G if it makes an angle 
with the neutron momentum smaller than 10 degrees. 

In the above discussion the mirror magnetic fields were neglected. 
However, one has to bear in mind that the mirror magnetic 
fields $B'$ may be actualy comparable with the ordinary ones, 
and therefore, the $n-n'$ mixing can be nearly maximal  
in the patches where the relative potential energy 
between $n$ and $n'$ states is small, 
$\Delta V = \gamma \mu | B_{\rm tr}  - B'_{\rm tr} | < \delta m $.  
On the other hand, it is likely that any cosmic ray faces in its            
propagation cosmic magnetic fields with varying intensities and orientations, 
and it is enough that the transverse component of 
${\bf B-B'}$ vanishes at some point to produce a resonant oscillation.  
If such accidental cancellations occur due to a spatial variation of the 
magnetic fields with a characteristic length larger than the oscillation 
length $c\, \tnn$, i.e.\ few kpc, 
then  adiabatic resonant transitions can take place at the borders 
between contiguous domains with different signs of $V-V'$, 
in which case 
the conversion probability is enhanced to $w=1$:
the Hamiltonian eigenstate being a dominantly $n$ state in one domain, 
can be almost fully converted into a dominantly $n'$ eigenstate 
at another domain, and  vice versa. 

One can also consider the possibility that the large scale ordinary and mirror 
magnetic fields coincide with a good precision. 
This situation could naturally emerge in the context of 
mechanisms of generation of magnetic seeds based on inflation scenarios 
in which the conformal invariance of quantum electrodynamics 
is broken due to a non-minimal coupling of the electromagnetic (EM) field 
with gravity or dilaton fields,  
or due to the quantum conformal anomaly for the trace 
of the EM field energy-momentum tensor~\cite{magn-inf}.
In this case, 
the couplings of the ordinary and mirror EM fields 
$F_{\mu\nu}$ and $F'_{\mu\nu}$ should be the same
due to mirror parity,\footnote{
In the context of models \cite{magn-inf} 
the relevant couplings could be e.g.  
$\cL \propto R_{\mu\nu\rho\sigma}
(F_{\mu\nu}F_{\rho\sigma} + F'_{\mu\nu}F'_{\rho\sigma})$ with the 
Riemann tensor $R_{\mu\nu\rho\sigma}$, or 
$\cL \propto (-1/4)\exp(S)(F_{\mu\nu}F^{\mu\nu} + 
F'_{\mu\nu}F^{\prime\mu\nu})$ with the dilaton field $S$.
} 
and thus identical magnetic seeds are generated 
in the O- and M- sectors. 
At the large scales of 100 Mpc, 
that still undergo a linear evolution, 
there can still remain a significant coincidence between the ordinary 
and mirror magnetic field values and their orientations, so that 
their compensation effect renders the effective $n-n'$ Hamiltonian  
nearly degenerate, $\Delta V  < \dm$. 
In this case the $n$ and $n'$ states have  
a nearly maximal mixing in voids, $w=1/2$.    
As for the non-linear scales, such a coherent cancellation 
cannot be expected because of the segregation 
between O- and M-matter components and consequently 
between the respective magnetic fields.


\section{Conclusions} 

The physics of such a familiar and long studied particle as the neutron 
still contains a big loophole:  
the experimental data do not exclude that its oscillation time into 
a mirror partner can be as small as 1 s~\cite{BB-nn}. 
This oscillation, however, is impossible for neutrons bound in nuclei or 
propagating in matter, and is suppressed by the terrestrial magnetic field, 
whereas it could easily occur for strictly free neutrons.   

The effect of such a fast $n-n'$ oscillation can be tested at small costs. 
"Table-top" experiments looking for the disappearance and/or regeneration of neutrons 
may reveal the neutron - mirror neutron oscillation,
with a good experimental control of the initial phase and mixing angle,
and thus open up a window to the mirror world. 
On the other hand, this phenomenon 
requires new physics at the TeV scale, with possible implications 
for the LHC~\cite{BB-nn,Nasri}, and, if mirror baryons contribute as 
dark matter of the Universe, it can have also  interesting astrophysical 
implications via the processes of mirror baryon scattering and/or annihilation 
with ordinary matter (depending on the sign of mirror baryon asymmetry). 

However, a fast $n-n'$ oscillation may have the most intriguing consequences 
for ultra high energy cosmic rays. 
Namely, it can enable UHE 
protons to propagate at large cosmological distances with no significant 
energy losses and thus explain the super-GZK excess in the 
cosmic ray spectrum as well as their directional correlation with 
far distant astrophysical sources. 

The complex approach and large statistics of the 
Pierre Auger experiment will permit to find out very soon 
if the anomalies in the ultra high energy cosmic ray spectrum is due 
to neutron - mirror neutron oscillation, or owing to other, 
as a matter of fact, not less exotic mechanisms as are e.g.\
the super-slow decay of super-heavy dark matter \cite{SHDM} 
or  Lorentz-violation~\cite{Lorentz}, 
while it is also possible that no super-GZK excess will be found at all.

\vspace{5mm}


\noindent {\bf Acknowledgments } 
\vspace{3mm} 

We are grateful to 
V. Berezinsky, A. Dolgov, S. Dubovsky, A. Gazizov, S.L. Glashow, 
S. Gninenko, O. Kancheli, V.\ Kuzmin, R.N.\ Mohapatra, V. Nesvizhevsky, 
S. Petrera, D.\ Semikoz, T.\ Soldner and I.\ Tkachev for illuminating 
discussions.
Z.B.\ was partially supported by the MIUR biennal grant for the Projects of 
the National Interest PRIN 2004 on "Astroparticle Physics".
L.B.\ was supported in part by the grant POCTI/FNU/43666/2002.

\newpage

\end{document}